\documentclass[12pt,a4paper]{article}
\usepackage{units}
\usepackage{graphicx}
\usepackage{amsmath,amssymb}
\DeclareMathOperator{\tr}{tr}
\DeclareMathOperator{\astcomma}{\stackrel{\ast}{,}}
\newcommand{\ii}{\mathrm{i}}
\newcommand{\ee}{\mathrm{e}}
\newcommand{\dd}{\mathrm{d}}
\allowdisplaybreaks

\makeatletter
\newcommand{\fmslash}[2][0mu]{%
  \mathchoice
    {\fmsl@sh\displaystyle{#1}{#2}}%
    {\fmsl@sh\textstyle{#1}{#2}}%
    {\fmsl@sh\scriptstyle{#1}{#2}}%
    {\fmsl@sh\scriptscriptstyle{#1}{#2}}}
\newcommand{\fmsl@sh}[3]{%
  \m@th\ooalign{$\hfil#1\mkern#2/\hfil$\crcr$#1#3$}}
\makeatother
\makeindex
\begin{document}
\title{%
  Probing the Noncommutative Standard Model at Hadron Colliders}
\author{%
  Ana Alboteanu%
    \thanks{e-mail: \texttt{aamaria@physik.uni-wuerzburg.de}}\qquad
  Thorsten Ohl%
    \thanks{e-mail: \texttt{ohl@physik.uni-wuerzburg.de}}\qquad
  Reinhold R\"uckl%
    \thanks{e-mail: \texttt{rueckl@physik.uni-wuerzburg.de}}\\
  \hfil\\
    Institut f\"ur Theoretische Physik und Astrophysik\\
    Universit\"at W\"urzburg, Am Hubland\\
    97074 W\"urzburg\\
    Germany}
\providecommand{\preprintno}[1]{\relax}
\preprintno{\hfil}
\date{%
  arXiv:hep-ph/0608155(rev)\\
  August 2006}
\maketitle
\begin{abstract}
  We study collider signals for the noncommutative extension of the
  standard model using the Seiberg-Witten maps for $\mathrm{SU}(3)_C\otimes
  \mathrm{SU}(2)_L\otimes \mathrm{U}(1)_Y$ to first order in the
  noncommutativity parameters
  $\theta_{\mu\nu}$.  In particular, we investigate the sensitivity of
  $Z\gamma$-production at the Tevatron and the LHC to the components of
  $\theta_{\mu\nu}$.  We discuss the range of validity of this
  approximation and estimate exclusion limits from
  a Monte Carlo simulation.
\end{abstract}
\newpage
\section{Introduction}
\label{sec:intro}

The observation that the low energy limit of certain open string
theories can be interpreted as Yang-Mills theories living on a
noncommutative~(NC)
manifold~\cite{Seiberg:1999vs} has raised renewed interest in
noncommutative quantum field theories.  Furthermore,
Noncommutative Quantum Field Theory~(NCQFT) in itself remains an
appealing way to introduce a length scale and to cut off some
short-distance contributions~\cite{Snyder} consistently with the
symmetries of a given model.

In the simplest case, the NC structure of spacetime
is described by a set of \emph{constant c-number}
parameters~$\theta^{\mu\nu}$, or, equivalently, by an energy scale
$\Lambda_{\text{NC}}$ and dimensionless parameters~$C^{\mu\nu}$:
\begin{equation}
\label{eq:theta}
  [\hat{x}^\mu, \hat{x}^\nu]
    = \ii \theta^{\mu\nu}
    = \ii \frac{1}{\Lambda_{\text{NC}}^2} C^{\mu\nu}
    = \ii \frac{1}{\Lambda_{\text{NC}}^2} 
            \begin{pmatrix}
                0 & - E^1 & - E^2 & - E^3 \\ 
              E^1 & 0     & - B^3 &   B^2 \\
              E^2 &   B^3 & 0     & - B^1 \\
              E^3 & - B^2 &   B^1 & 0
            \end{pmatrix}\,.
\end{equation}
Here we have defined dimensionless ``electric''
and ``magnetic''
parameters~$\vec E$ and~$\vec B$ for convenience.  Obviously,
fixed non-vanishing values for~$\theta^{\mu\nu}$ result in a breaking
of Lorentz invariance.  A possible mechanism of spontaneous breaking
is the condensation of a rank two antisymmetric background field.  In
this paper, we will investigate phenomenological consequences
of~$\theta^{\mu\nu}$ at hadron colliders and not deal with
the dynamics of the symmetry breaking mechanism.

Since the string scale is essentially unknown, we have also no prediction for
$\Lambda_{\text{NC}}$.  The current best
collider tests~\cite{Abbiendi:2003wv} give
$\Lambda_{\text{NC}}>\unit[140]{GeV}$ for a NC extension of QED, while
experiments at a future photon collider are expected to reach
\unit[1]{TeV}~\cite{Ohl/Reuter:2004:NCPC}.  Tests
of rotation invariance in atomic physics and
astrophysical tests give substantially higher bounds, but assume NC geometry
at large scales (see~\cite{Hinchliffe:2002km} for a review).

Collider experiments do not probe space-time coordinates themselves,
but functions of space-time coordinates, such as wave functions of
external particles and effective Lagrangians.  Consequently, one can
employ a deformed product to describe the scattering theory on a given
NC manifold by an equivalent scattering theory on a commuting manifold
with new effective interactions.  In the case of the operator
algebra~(\ref{eq:theta}), one can use the Moyal-Weyl
$\ast$-product~\cite{Moyal:1949sk}
 \begin{equation}
   \label{eq:moyal-weyl}
   (f\ast g)(x) = 
   f(x) e^{ \frac{\ii}{2}\overleftarrow{\partial_\mu}
     \theta^{\mu\nu}
     \overrightarrow{\partial_\nu}} g(x)
     = f(x)g(x) + \frac{\ii}{2} \theta^{\mu\nu}
     \frac{\partial f(x)}{\partial x^\mu}
     \frac{\partial g(x)}{\partial x^\nu}
     + \mathcal{O}(\theta^2)
 \end{equation}
to map products of functions of NC coordinates, $(fg)(\hat x) =
f(\hat x)g(\hat x)$, to the equivalent products of functions on
ordinary space-time coordinates, $(f\ast g)(x)$.  It is easy to see that
the commutation relation~(\ref{eq:theta}) is reproduced by the
$\ast$-commutator:
 \begin{equation}
   [x^\mu \astcomma x^\nu](x)
   =   (x^\mu \ast x^\nu)(x)
   - (x^\nu \ast x^\mu)(x)
   = \ii \theta^{\mu\nu}=[\hat{x}^\mu,\hat{x}^\nu]\,.
 \end{equation}
The obvious approach to constructing NCQFTs by replacing all field
products by the $\ast$-product~(\ref{eq:moyal-weyl}) fails for
general gauge theories.  For example, in such a NC version of QED all
charged matter field must carry the same charge~\cite{Hayakawa:1999yt}
which makes this approach unsuitable for the $\mathrm{U}(1)$
hypercharge of the standard
model. Furthermore, this approach does not work for
$\mathrm{SU}(N)$ gauge theories, because the $\ast$-commutator of two
infinitesimal gauge transformations does not close in the
$\mathrm{SU}(N)$ Lie algebra~\cite{Wess:2001pr}:
\begin{equation}
  [\alpha \astcomma \beta] =
  [\alpha^aT^a \astcomma \beta^bT^b] =
  \frac{1}{2}\{\alpha^a\astcomma\beta^b\}[T^a,T^b] 
  + \frac{1}{2}[\alpha^a\astcomma\beta^b]\{T^a,T^b\} \,.
\end{equation}
In order to circumvent these problems, we follow the approach
of~\cite{Wess:2001pr} and let the NC gauge field~$\hat A$ and gauge
parameter~$\hat\lambda$ take their values in the enveloping algebra
of the Lie algebra of the Standard Model~(SM) gauge group
$\mathrm{SU}(3)_C\otimes \mathrm{SU}(2)_L\otimes \mathrm{U}(1)_Y$.
$\hat A$, $\hat\lambda$ and the NC matter fields~$\hat\psi$ are
expressed as nonlinear functions of ordinary fields $A$, $\lambda$ and~$\psi$
via Seiberg-Witten maps (SWM)~\cite{Seiberg:1999vs} such that the NC gauge
transformations of the former,
\begin{subequations}
\label{eq:NC-gauge-transformations}
\begin{align}
  \hat\psi &\rightarrow \hat\psi'
     = \ee^{\ii g\hat\lambda\ast} \hat\psi
     = \hat\psi + \ii g\hat\lambda\ast \hat\psi
            + \frac{(\ii g)^2}{2!}
                \hat\lambda\ast\hat\lambda\ast\hat\psi
            + \mathcal{O}(\hat\lambda^3) \\
  \hat A_\mu &\rightarrow \hat A_\mu' =
    \ee^{\ii g\hat\lambda\ast}
       \hat A_\mu \ee^{- \ii g\hat\lambda\ast} 
      + \frac{\ii}{g}\ee^{\ii g\hat\lambda\ast}
           \left(\partial_\mu \ee^{- \ii g\hat\lambda\ast} \right) \\
        &\phantom{\rightarrow \hat A_\mu'}
  = \hat A_\mu + \ii g [\hat\lambda\astcomma\hat A_\mu]
          + \frac{(\ii g)^2}{2!} [\hat\lambda\astcomma[\hat\lambda\astcomma\hat A_\mu]]
          + \partial_\mu \hat\lambda
          + \ii g [\hat\lambda\astcomma\partial_\mu \hat\lambda]
          + \mathcal{O}(\hat\lambda^3)\,, \notag
\end{align}
\end{subequations}
are realized by ordinary gauge transformations of the latter:
\begin{subequations}
\label{eq:gauge-transformations}
\begin{align}
  \psi &\rightarrow \psi'
     = \ee^{\ii g\lambda} \psi
     = \psi + \ii g\lambda \psi
            + \frac{(\ii g)^2}{2!}
                \lambda^2\psi
            + \mathcal{O}(\lambda^3) \\
  A_\mu &\rightarrow A_\mu' =
    \ee^{\ii g\lambda}
       A_\mu \ee^{- \ii g\lambda} 
      + \frac{\ii}{g}\ee^{\ii g\lambda}
           \left(\partial_\mu \ee^{- \ii g\lambda} \right) \\
        &\phantom{\rightarrow A_\mu'}
  = A_\mu + \ii g [\lambda,A_\mu]
          + \frac{(\ii g)^2}{2!} [\lambda, [\lambda, A_\mu]]
          + \partial_\mu \lambda
          + \ii g [\lambda,\partial_\mu \lambda]
          + \mathcal{O}(\lambda^3)\,. \notag
\end{align}
\end{subequations}
This leads to the so-called gauge-equivalence conditions
\begin{subequations}
\label{eq:SWM-condition}
\begin{align}
  \hat A (A, \theta)
     &\to \hat A' (A, \theta) = \hat A (A', \theta) \\
  \hat \lambda (\lambda, A, \theta)
     &\to \hat \lambda' (\lambda, A, \theta) = \hat \lambda (\lambda', A', \theta) \\
  \hat \psi (\psi, A, \theta)
     &\to \hat \psi' (\psi, A, \theta) = \hat \psi (\psi', A', \theta)\,,
\end{align}
\end{subequations}
that can be
solved order by order in~$\theta^{\mu\nu}$. The first order is given
by
\begin{subequations}
\begin{align}
  \hat A_\mu(x) &= A_\mu(x)
      + \frac{1}{4} \theta^{\rho\sigma}
                    \left\{A_\sigma(x),
                          \partial_\rho A_\mu(x) + F_{\rho\mu}(x)\right\}
      + \mathcal{O}(\theta^2) \\
  \hat \psi (x) &= \psi(x)
      + \frac{1}{2} \theta^{\rho\sigma}
                    A_\sigma(x) \partial_\rho \psi(x)
      + \frac{\ii}{8} \theta^{\rho\sigma}
                      \left[A_\rho(x), A_\sigma(x)\right] \psi(x)
       + \mathcal{O}(\theta^2)\\
  \hat \lambda(x) &= \lambda(x)
      + \frac{1}{4} \theta^{\rho\sigma}
                    \left\{A_\sigma(x), \partial_\rho \lambda(x)\right\}
       + \mathcal{O}(\theta^2)
\end{align}
\end{subequations}
with $F_{\mu\nu}=(\ii/g)[D_\mu,D_\nu]$.
The SWM approach also solves the charge quantization problem by
introducing a
separate NC gauge field  for each eigenvalue of the
$\mathrm{U}(1)$ charge operator.  Each of these gauge fields is
subsequently mapped by the SWM to the \emph{same} SM
$\mathrm{U}(1)$ field.  Thus no additional degrees of freedom appear.
This approach has been used to construct a NC generalization of the
complete electroweak standard model~\cite{NCSM}, which will be
referred to as NCSM below.

This construction leads to modifications of the familiar SM interaction
vertices as well as to new interactions that are forbidden in the SM.
The most striking feature of the NCSM is that it allows direct interactions
among neutral gauge bosons, depending on the representation of the
enveloping algebra chosen for the trace in the gauge boson Lagrangian~\cite{NCSM}
\begin{equation}
\label{eq:tr(F^2)}
  \mathcal{L} = - \frac{1}{2} \tr\Bigl( F_{\mu\nu}F^{\mu\nu} \Bigr)\,.
\end{equation}
Note that all representations are equivalent up to a
renormalization only in the case of Lie algebras, but not in the
corresponding enveloping associative algebra.

We should point out that NC gauge theories constructed from
commutative gauge theories using the SWM are not renormalizable by
field and gauge coupling renormalization alone, since the $4$-point
functions of matter fields require additional counterterms that can
not be obtained from the application of the
SWM~\cite{Renormalization}.  Nevertheless, the NCSM can be employed as
a well defined effective field theory, because any chiral model that
is constructed from an anomaly free chiral model using the SWM has
been shown to be anomaly free~\cite{Anomalies}.

To first order in~$\theta^{\mu\nu}$, all modifications of the SM
correspond to effective dimension-six operators.  The latter lead to
observable deviations from the SM predictions or, in the absence of
deviations, to bounds on $\Lambda_{\text{NC}}$.  Initially, most
phenomenological studies have focused on rare and
$B$-decays~\cite{Collected-Munich/Zagreb-Works}, while tests at future
colliders have been considered in~\cite{Ohl/Reuter:2004:NCPC}.
Additional phenomenological results have been presented in~\cite{More-Pheno}.

In this paper, we report on a phenomenological study of the NCSM in
first order in~$\theta^{\mu\nu}$, concentrating on the process
$pp/p\bar p\to \ell^+\ell^- \gamma$
at hadron colliders. We shall point out experimental
signatures unique to the NCSM and estimate the range of the NC
scale~$\Lambda_{\text{NC}}$ that can be probed at the Tevatron and the
LHC.  The paper is structured as follows: In section~\ref{sec:xsect},
we discuss the scattering amplitude that gives rise to the signals described in
section~\ref{sec:signals}.  In section~\ref{sec:analysis}, we analyse
the results of our Monte Carlo~(MC) simulations and estimate the
discovery reach. Section~\ref{sec:concl} gives a short outlook on
future work. Feynman rules and analytical results for
the cross section are presented in the appendix.

\section{Partonic Scattering Amplitudes and Cross Sections}
\label{sec:xsect}

\begin{figure}
  \begin{center}
    \includegraphics{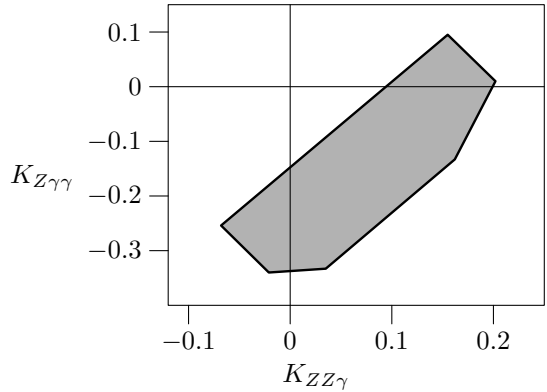}
  \end{center}
  \caption{\label{fig:K_Zgg/K_ZZg}%
    Range for the triple neutral gauge boson couplings
    $K_{Z\gamma\gamma}$ and $K_{ZZ\gamma}$ allowed by the matching to
    the SM in the limit $\theta^{\mu\nu}\to0$ in the nonminimal
    NCSM, as defined in~\cite{NCSM}.}
\end{figure}

The pair production of neutral electroweak gauge bosons $V=\gamma,Z$
at a hadron collider is described by the partonic process $q\bar q\to
VV'$.  The corresponding squared matrix element at
$\mathcal{O}(\theta)$ is given by
\begin{equation}
\label{eq:|A|^2}
  |A|^2 = |A^{\text{SM}}+A^{\text{NC}}|^2
     = |A^{\text{SM}}|^2
     + 2\mathrm{Re} \left(A^{*\,\text{SM}} A^{\text{NC}}\right)
     + O(\theta^2)
\end{equation}
where $A^{\text{SM}} = A^{\text{SM}}_t + A^{\text{SM}}_u$ is the SM
amplitude with
\begin{equation}
  A^{\text{SM}}_t =
    \parbox{36mm}{\includegraphics{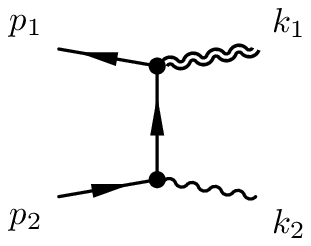}} \qquad
  A^{\text{SM}}_u =
    \parbox{36mm}{\includegraphics{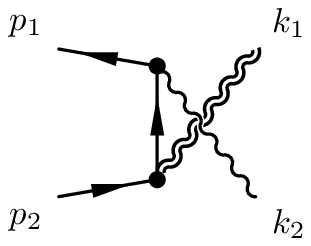}}
\end{equation}
and
\begin{equation}
   A^{\text{NC}}
    = \sum_{i=1,2} (A^{\text{NC}}_{t,i} + A^{\text{NC}}_{u,i})
       + A^{\text{NC}}_{c} + A^{\text{NC}}_{s,\gamma} + A^{\text{NC}}_{s,Z}\,,
\end{equation}
denotes the new contributions arising in the NCSM.  The relevant
Feynman rules are provided in appendix~\ref{app:feynman}.  Below,
the $\mathcal{O}(\theta)$ contributions to the SM
vertices~(\ref{eq:vertex-ffV}) have been marked by an open box in the
corresponding SM Feynman diagrams:
\begin{subequations}
\label{eq:NC-amplitude}
\begin{align}
\label{eq:NC-t-channel}
  A^{\text{NC}}_{t,1} &=
    \parbox{36mm}{\includegraphics{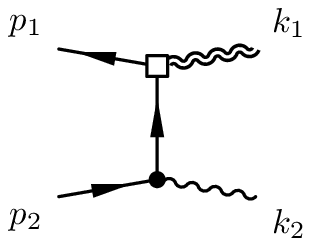}} &\qquad&
  A^{\text{NC}}_{t,2} &=
    \parbox{36mm}{\includegraphics{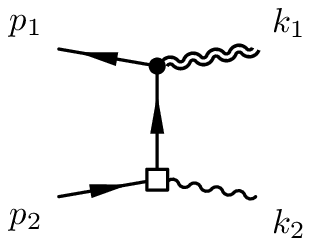}} \\
\label{eq:NC-u-channel}
  A^{\text{NC}}_{u,1} &=
    \parbox{36mm}{\includegraphics{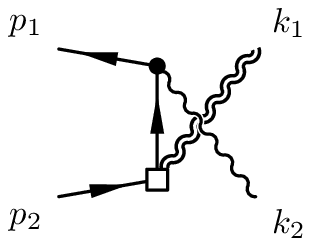}} &\qquad&
  A^{\text{NC}}_{u,2} &=
    \parbox{36mm}{\includegraphics{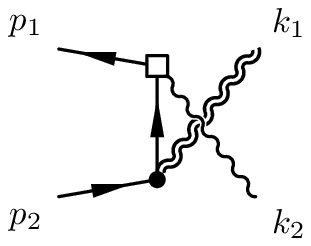}}\,.
\end{align}
The contact term
\begin{equation}
\label{eq:NC-contact}
  A^{\text{NC}}_c =
    \parbox{36mm}{\includegraphics{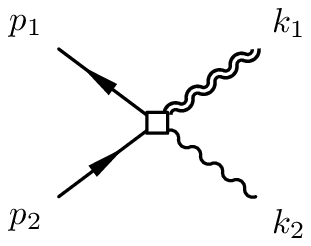}}\,,
\end{equation}
with the vertex given in~(\ref{eq:vertex-ffVV}), is required by gauge
invariance, as can be seen from the Ward identity~(\ref{eq:WIa}).  In
contrast, the two $s$-channel diagrams
\begin{equation}
\label{eq:NC-s-channel}
  A^{\text{NC}}_{s,\gamma} =
    \parbox{36mm}{\includegraphics{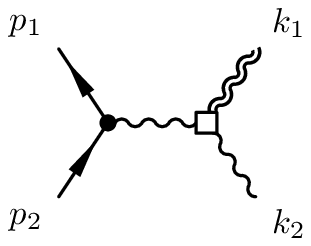}} \qquad
  A^{\text{NC}}_{s,Z} =
    \parbox{36mm}{\includegraphics{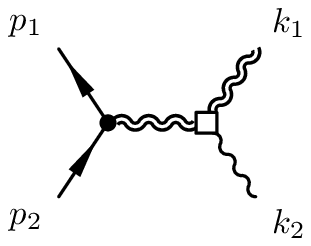}}
\end{equation}
\end{subequations}
involving the triple gauge boson couplings given
in~(\ref{eq:vertex-VVV}) are separately gauge
invariant.  The corresponding Ward identities
\begin{subequations}
\begin{align}
\label{eq:WIa}
  k_{1,\mu} \Bigl(
     \sum_{i=1}^2 (A^{\text{NC}}_{t,i} + A^{\text{NC}}_{u,i})
        + A^{\text{NC}}_{c}
  \Bigr)^{\mu\nu} \varepsilon_\nu(k_2) &= 0 \\
\label{eq:WIb}
  k_{1\mu}( A^{\text{NC}}_{s,\gamma})^{\mu\nu}\varepsilon_\nu(k_2) =
  k_{1\mu}( A^{\text{NC}}_{s,Z})^{\mu\nu}\varepsilon_\nu(k_2) &= 0\,.
\end{align}
\end{subequations}
have been proven in~\cite{Ohl/Reuter:2004:NCPC}.
The separate gauge invariance of~$A^{\text{NC}}_{s,\gamma}$
and~$A^{\text{NC}}_{s,Z}$ is not surprising, because the strength of
the triple gauge boson interactions depends on the choice of
representation of the enveloping algebra~\cite{NCSM} used for the gauge field
Lagrangian~(\ref{eq:tr(F^2)}). The new coupling
constants~$K_{Z\gamma\gamma}$ and~$K_{ZZ\gamma}$
are only constrained by the matching to the SM in the limit
$\theta^{\mu\nu}\to0$ and can vary independently in a finite
range~\cite{NCSM}, as illustrated in fig.~\ref{fig:K_Zgg/K_ZZg}.

It has been shown~\cite{Ohl/Reuter:2004:NCPC} for
$\gamma\gamma$-collisions that polarization is essential for obtaining
a non-vanishing signal in $\mathcal{O}(\theta)$. At
the LHC one will not have polarized initial states,
neither can one expect to observe the polarization of
outgoing photons or use $Z$-decays as a polarimeter.  Therefore, we have
concentrated on unpolarized $Z\gamma$-production.
The analytical result for $\dd\sigma(q\bar q\to Z\gamma)/\dd\Omega$ is
presented in appendix~\ref{app:xsect}.  

Without polarization, the most important observable, i.\,e.~an
azimuthal dependence of the cross section, vanishes for
symmetric initial and final states, as we will see in
section~\ref{sec:signals}. In the present case, it is the axial coupling of
the~$Z$ that will engender an observable deviation from the SM
prediction.  Since the width of the intermediate $Z$ is
negligible for $\sqrt{\hat s} \gg m_Z$, one needs
another source for an imaginary part in the scattering
amplitudes to compensate the factor of $\ii$ that comes with
$\theta^{\mu\nu}$ and to obtain a non-vanishing contribution to the
$\mathcal{O}(\theta)$ interference.  In polarized scattering, the
polarization vectors and spinors can provide this imaginary part, but
all the polarization sums are real, of course.  However, the
$\gamma^5$ factor in the axial coupling of the $Z$ produces a factor
$\ii\epsilon^{\mu\nu\rho\sigma}$ after tracing and thus creates a
non-vanishing effect from contractions with the three independent
momenta and~$\theta^{\mu\nu}$.

\section{Distributions and Signals in $pp/p\bar p\to Z\gamma$}
\label{sec:signals}

The relevant hadronic processes are $pp\to Z\gamma$ at the LHC and
$p\bar p\to Z\gamma$ at the Tevatron with subsequent decays of the $Z$
into $e^+e^-$ and $\mu^+\mu^-$.  For these, the cross sections and
lepton distributions have been calculated using the
results for the hard $q\bar q \to \ell^+\ell^-\gamma$ cross sections
from section~\ref{sec:xsect} together with the phase space generation
and parton distribution functions provided by
WHiZard~\cite{Kilian:WHIZARD} for Monte Carlo simulation.

As expected for new short distance physics, we observe a deviation of the
photon $p_T$ distribution at high $p_T$ from the SM prediction.
However, this deviation is not specific to the NCSM and, in addition,
turns out to be smaller than a few~$\%$
for~$p_T\lesssim\Lambda_{\text{NC}}/2$. On the other hand, the
NC parameter $\theta^{\mu\nu}$ breaks rotational
invariance and leads to a characteristic dependence of the cross
sections on the azimuthal angle~$\phi$, which can be used to
discriminate against other new physics effects.  For this reason, we
have focussed our phenomenological analysis on the azimuthal
distribution of the photon.

\begin{figure}
  \begin{center}
    \includegraphics{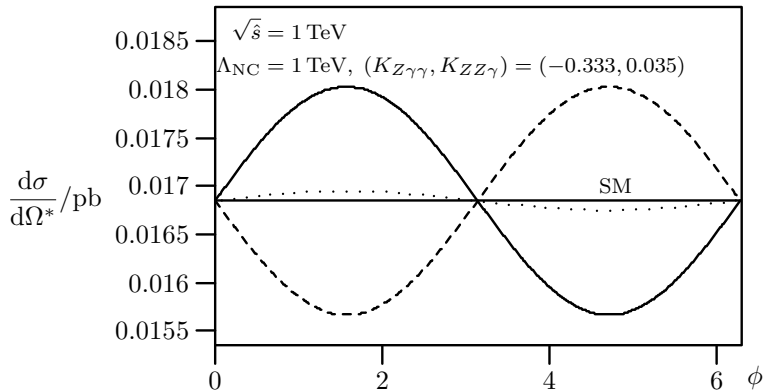}
  \end{center}
  \caption{\label{fig:azimuthal/E/B}%
    Azimuthal distribution of the $\gamma$ in $q\bar q\to Z\gamma$ at
    fixed CMS energy $\sqrt{\hat s}$ for
    $\vec{E} = (1,0,0)$ and $\cos\theta^* = 0.05 \text{(full)}, -0.05
    \text{(dashed)}$, in comparison with $\vec{B} = (0,1,0)$ for
    $\cos\theta^* = \pm0.05 \text{(dotted)}$.}
\end{figure}

As has been shown previously~\cite{Ohl/Reuter:2004:NCPC}, the
$\gamma\gamma\to f\bar f$ amplitude in the NCSM depends only on $E_1$
and $E_2$.  In contrast, the $f\bar f\to
Z\gamma$ amplitude also depends on~$B_1$ and~$B_2$,
due to the axial $Zf\bar f$ couplings. In addition, a
dependence on~$E_3$ appears because the vector bosons are not aligned
with the beam axis~$x_3$.  Nevertheless, the CMS parton
cross sections show a much stronger dependence on the
components $\vec E$ than on $\vec B$, as exemplified in
fig.~\ref{fig:azimuthal/E/B}.
To be precise, this is true everywhere except sufficiently close to
the polar angle $\theta^*=\pi/2$, where the dependence
on $\vec E$ vanishes due to the antisymmetry of the
$\mathcal{O}(\theta)$ contribution to
$\dd\sigma/\dd\Omega^*$ in $\cos\theta^*$.

However, at hadron colliders, the partonic CMS of most events is
boosted significantly along the beam axis\footnote{The boosted events
are further enriched by the cuts chosen for the LHC below.}.  As is
well known from electrodynamics, $\vec E$ and $\vec B$ are mixed by
Lorentz boosts along the beam axis $x_3$:
\begin{equation}
\label{eq:correlation}
  \begin{aligned}
     E_1 &\to \gamma(E_1 - \beta B_2) &\quad B_1 &\to \gamma(B_1 + \beta E_2) \\
     E_2 &\to \gamma(E_2 + \beta B_1) &\quad B_2 &\to \gamma(B_2 - \beta E_1) \\
     E_3 &\to E_3                 &\quad B_3 &\to B_3
  \end{aligned}
\end{equation}
($\gamma= 1/\sqrt{1-\beta^2}$, $\beta = v/c$).  The measurements of
$(E_1,B_2)$ and $(E_2,B_1)$, respectively, are therefore highly correlated by
kinematics.  We have verified that the strenght of the correlation is essentially
determined by the expectation value of the boost $\langle |\beta| \rangle$.

\label{sec:unphysical}

Calculating the scattering amplitude only to first order in the
noncommutativity~$\theta^{\mu\nu}$, one must neglect the
$\mathcal{O}(\theta^2)$ contributions in the squared amplitude for
consistency.  Therefore, in this approximation, the deviations from
the SM only come from the interference of~$A^{\text{SM}}$
with~$A^{\text{NC}}$ in~(\ref{eq:|A|^2}).  As a consequence, the
$\mathcal{O}(\theta)$ cross section can even become negative in some
regions of phase space for sufficiently large absolute values and
appropriate signs of components of~$\theta^{\mu\nu}$, such that the NC/SM
interference dominates over the SM contribution.
The relative size of the interference term is determined by $\hat
s/\Lambda_{\text{NC}}^2$, $\sqrt{\hat s}$ being the partonic CMS energy.
There is a wide range of possible values of $\sqrt{\hat s}$ in
high energy hadron collisions, but the most statistics will be
collected at moderate values of~$\sqrt{\hat s}$.  Thus in
$\mathcal{O}(\theta)$, values of~$\Lambda_{\text{NC}}$ that cause
observable deviations at such moderate values of~$\sqrt{\hat s}$ can
lead to unphysical cross sections at the highest~$\sqrt{\hat s}$
available.
This problem is not specific to simulations in the NCSM, but common to
all studies of new physics that can be parametrized by anomalous
couplings.  A pragmatic solution is to unitarize the contributions from new
physics by applying appropriate form factors that cut off the unphysical
effects. Since
there are very few events to be expected at the highest CMS energies, the
conclusions should not depend on the details of the form factors.
Therefore, we have simply replaced $\dd\sigma/\dd\Omega^*$ by
$\max(\dd\sigma/\dd\Omega^*,0)$ everywhere in our simulations.

This solution is also legitimized by the expectation that higher
orders in~$\theta^{\mu\nu}$ will damp the large negative interference
contributions.  In fact, preliminary results to
second order in $\theta^{\mu\nu}$ support this
expectation~\cite{Alboteanu/Ohl/Rueckl:2005:theta2}.

\section{Sensitivity Estimates}
\label{sec:analysis}

In order to obtain a realistic estimate of the sensitivity at the
Tevatron and the LHC, one has to take into account backgrounds,
detector effects and selection cuts.  Clearly, a comprehensive
analysis of all reducible backgrounds and detector effects is beyond
the scope of a theoretical study and must eventually be performed by
the experimental collaborations.  However, the final states under
consideration are simple enough for a phenomenological analysis based
on simple cuts.  Moreover, experience at the Tevatron~\cite{CDF,D0}
indicates that the combined detection efficiency for
$\ell^+\ell^-\gamma$ can be assumed to be larger than $50\%$.  All
numerical results presented below are obtained for a $100\%$
efficiency.
Smaller uniform efficiencies can easily be taken into account by
scaling up the integrated luminosity accordingly.

We have simulated the process $pp\to\ell^+\ell^-\gamma$ at
$\sqrt{s}=\unit[14]{TeV}$ using the program package
WHiZard~\cite{Kilian:WHIZARD} and  demanding the following acceptance
cuts on leptons and photons:
\begin{subequations}
\label{eq:acceptance}
\begin{align}
  E(\ell^\pm), E(\gamma)           &\ge \unit[10]{GeV} \\
  \theta(\ell^\pm), \theta(\gamma) &\ge 5^\circ \\
  \theta(\ell^\pm), \theta(\gamma) &\le 175^\circ \\
  p_T(\ell^\pm), p_T(\gamma)       &\ge \unit[10]{GeV}\,.
\end{align}
\end{subequations}
For the Tevatron, we have simulated $p\bar p\to\ell^+\ell^-\gamma$ at
$\sqrt{s}=\unit[1.96]{TeV}$ using the same acceptance
cuts~(\ref{eq:acceptance}).

For simplicity, we have assumed that the components of
$\theta^{\mu\nu}$ remain aligned with the beam axis and the detector
over the time of the measurement.  This assumption is not justified,
because we should expect that $\theta^{\mu\nu}$ is aligned with a
fixed cosmic reference frame, that is determined by the dynamics of
the underlying string theory.  Therefore the alignment of the detector
must be recorded with each event and the combined effect of the
earth's rotation and revolution must be taken into account. This poses
no principal difficulty and will not change our conclusions.

\subsection{Background Suppression}
The hard scattering process $q\bar q\to Z\gamma$ (\ref{eq:NC-amplitude}) with
subsequent decays $Z\to\ell^+\ell^-$ has the Drell-Yan process $q\bar
q\to \ell^+\ell^-$ with photon radiation and $q\bar q\to
\gamma^*\gamma$ with $\gamma^*\to\ell^+\ell^-$ as irreducible
backgrounds.  Both have been taken into account in our calculation.

As explained in section~\ref{sec:xsect}, the unpolarized azimuthal
distribution in $q\bar q\to \gamma^*\gamma$ is flat. We
suppress this background by requiring
\begin{equation}
  |M(\ell^+\ell^-)-M_Z| \leq \Gamma_Z\,.
\end{equation}
In order to also reduce the
radiative Drell-Yan events, we have applied an angular separation
cut of
\begin{equation}
  \Delta R_{\ell^{\pm}\gamma} = \sqrt{\Delta \eta ^2 + \Delta \phi^2} > 0.7
\end{equation}
(cf.~\cite{CDF,D0}).  Finally, we require a minimum and maximum total
energy in the partonic CMS:
\begin{equation}
  \unit[200]{GeV} \leq |M(\ell^+\ell^-\gamma)| \leq \unit[1]{TeV}\,.
\end{equation}
The lower cut enriches the signal, while the upper cut
reduces the influence of the elimination of the unphysical parton cross
sections discussed in section~\ref{sec:unphysical}.

At the LHC, $q\bar q$- and $\bar q q$-collisions
will occur with identical rates.  As shown in
fig.~\ref{fig:azimuthal/E/B}, the dominant $E_1$ and $E_2$
contributions to the deviation of the parton cross sections from the SM
prediction is antisymmetric in $\cos{\theta^*}$ and will
therefore cancel for $pp$ initial states unless additional cuts are
applied that separate events originating from $q\bar{q}$ and $\bar{q}q$.

\begin{figure}
  \begin{center}
    \includegraphics{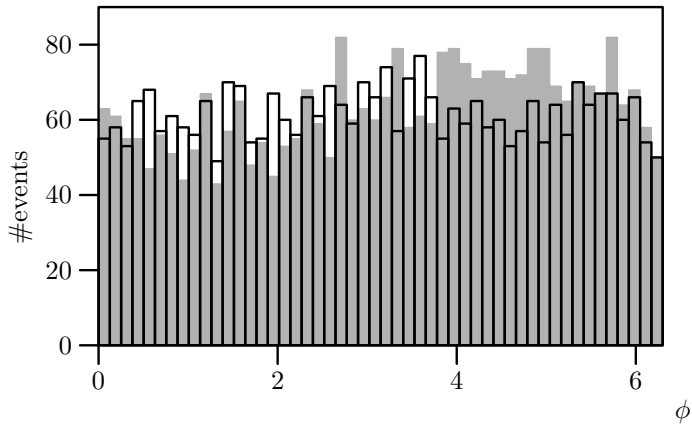}
  \end{center}
  \caption{\label{fig:azimuthal}%
    Azimuthal distributions of the~$\gamma$ in $pp\to e^+e^-\gamma$
    at the LHC, $\sqrt{s} = 14\,\text{TeV}$, $300\,\text{fb}^{-1}$,
    for $\cos\theta_{e^+e^-} > 0$, $\cos\theta_\gamma > 0$, 
    $0<\cos\theta^*_\gamma<0.9$ in the SM (open) and the NCSM (full)
    with $\Lambda_{\text{NC}} = 0.6\,\text{TeV}$,
    $\vec E = (1,0,0)$, $K_{Z\gamma\gamma}=0.095$, $K_{ZZ\gamma}=0.155$.}
\end{figure}

In the proton, the average momentum fraction of the light valence
quarks is much higher than that of the anti-quarks which exist only
in the sea.  As a result, all $q\bar q$-events will be boosted
strongly in the direction of the quark.  Therefore we can enrich our
samples of signal events by requiring a minimal
boost in the appropriate direction.  We have found that demanding
the momenta of both the photon and the lepton pair to lie in the
\emph{same} hemisphere in the laboratory frame, that is
\begin{equation}
  \cos{\theta_\gamma} \cdot \cos{\theta_{\ell^+\ell^-}} > 0
\end{equation}
produces a nice signal, as displayed in fig.~\ref{fig:azimuthal}.  In
the figure, we have chosen a rather low value of
$\Lambda_{\text{NC}} = 0.6\,\text{TeV}$ for illustration.  The values of
$K_{Z\gamma\gamma}$ and $K_{ZZ\gamma}$ correspond to the top corner of
the polygon in fig.~\ref{fig:K_Zgg/K_ZZg}.

In the case of the Tevatron, the r\^ole of quarks and anti-quarks are
reversed in the antiproton.  Therefore, we demand in this case that
the momenta of the photon and the lepton pair lie in
\emph{opposite} hemispheres.  For sufficiently small
$\Lambda_{\text{NC}}$, the azimuthal distribution of the~$\gamma$ is
similar to the distribution plotted in fig.~\ref{fig:azimuthal}.

\subsection{Likelihood Fits}
In the following, we analyze the NC effects on the azimuthal
dependence of the unpolarized cross section stemming from
the components $E_1, E_2, B_1, B_2$ of~$\theta_{\mu\nu}$.  We
disregard the small contribution from $E_3$, since, by symmetry, it
has no influence on the azimuthal distribution.

Furthermore, we have taken advantage of the fact that the
deviation of the cross section from the SM prediction is
linear in~$\theta_{\mu\nu}$ so that the corresponding binned $\chi^2$-function
\begin{equation}
\label{eq:chi^2}
  \chi^2(E_1, E_2, B_1, B_2)
    = \sum_i \frac{1}{n^{\text{SM}}_i} \left(n^{\text{NC}}_i-n^{\text{SM}}_i\right)^2
\end{equation}
depends quadratically on~$\theta_{\mu\nu}$:
\begin{equation}
\label{eq:chi^2/X^2}
  \chi^2(E_1, E_2, B_1, B_2)
    = \frac{1}{2}
         (E_1, E_2, B_1, B_2) X^2
         (E_1, E_2, B_1, B_2)^T\,,
\end{equation}
up to statistical fluctuations.  The elements of the matrix~$X^2$
in~(\ref{eq:chi^2/X^2}) are then determined from a fit of~$\chi^2(E_1,
E_2, B_1, B_2)$ to the right-hand side of~(\ref{eq:chi^2})
for a sufficiently large grid of values~$(E_1, E_2, B_1, B_2)$.
Subsequently, we can diagonalize $X^2$ and obtain the
error ellipses that describe the experimental reach of collider
experiments.

\subsection{Results}

\begin{figure}
  \begin{center}
    \includegraphics[width=0.30\textwidth]{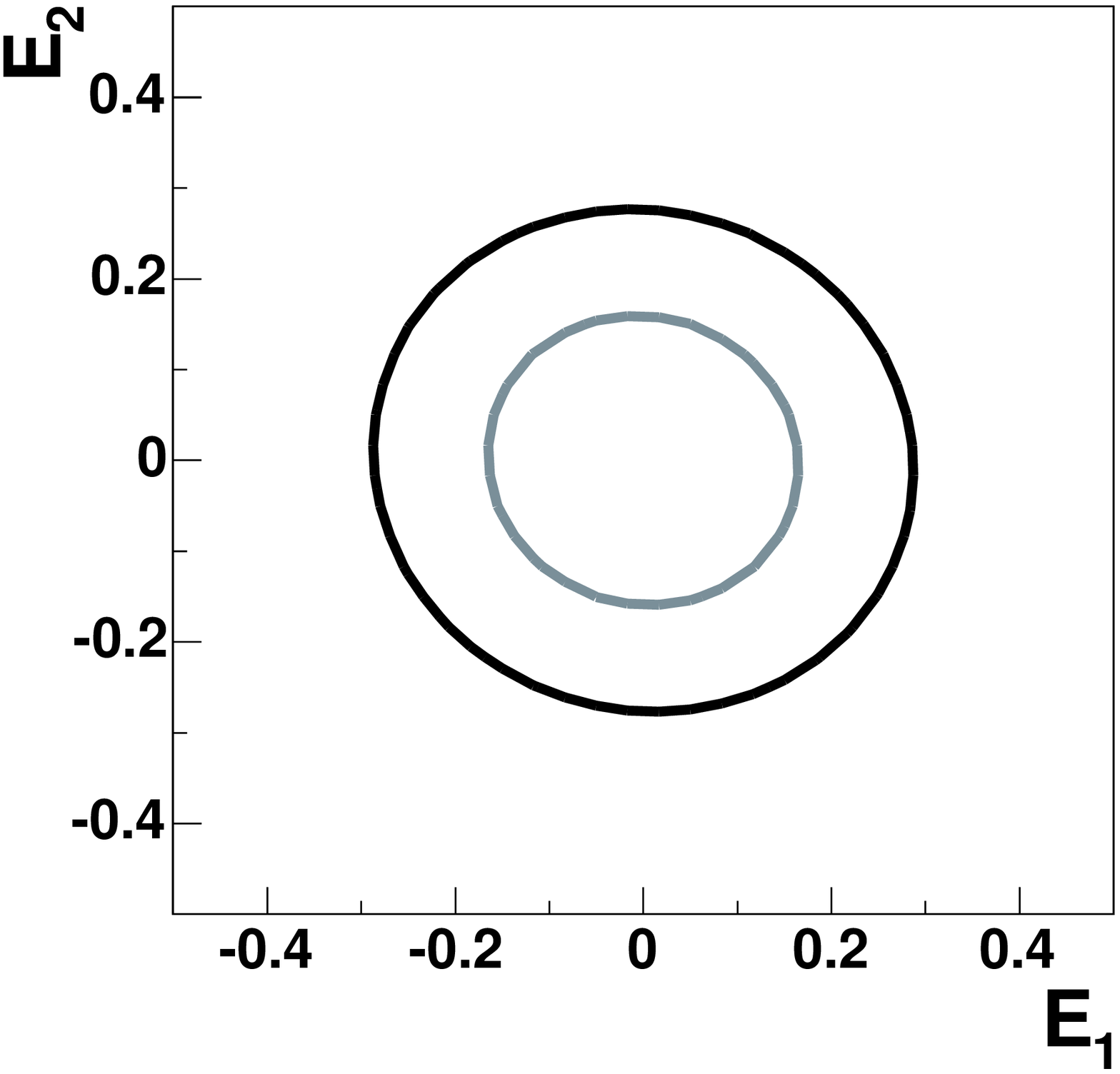}\quad
    \includegraphics[width=0.30\textwidth]{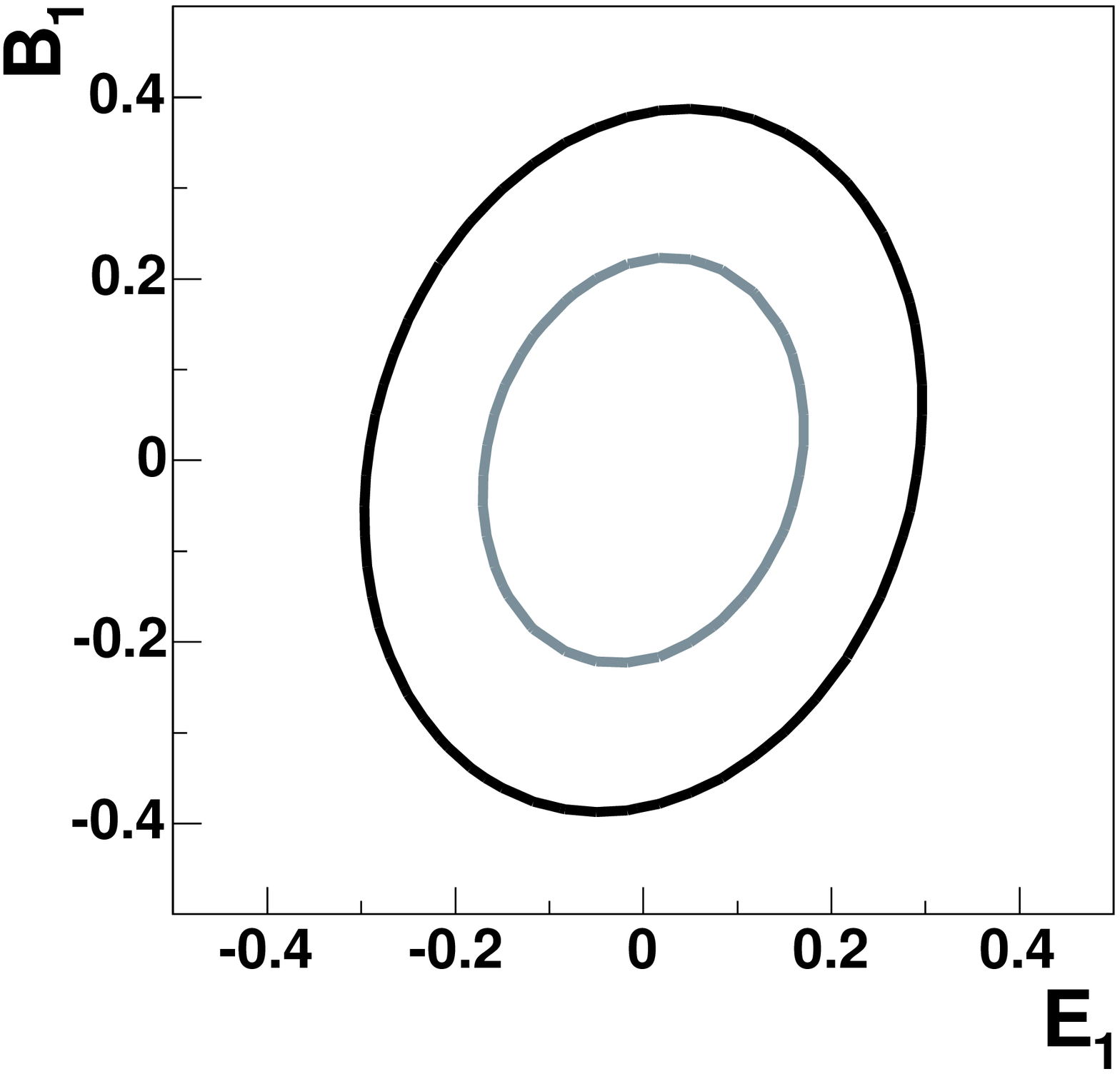}\quad
    \includegraphics[width=0.30\textwidth]{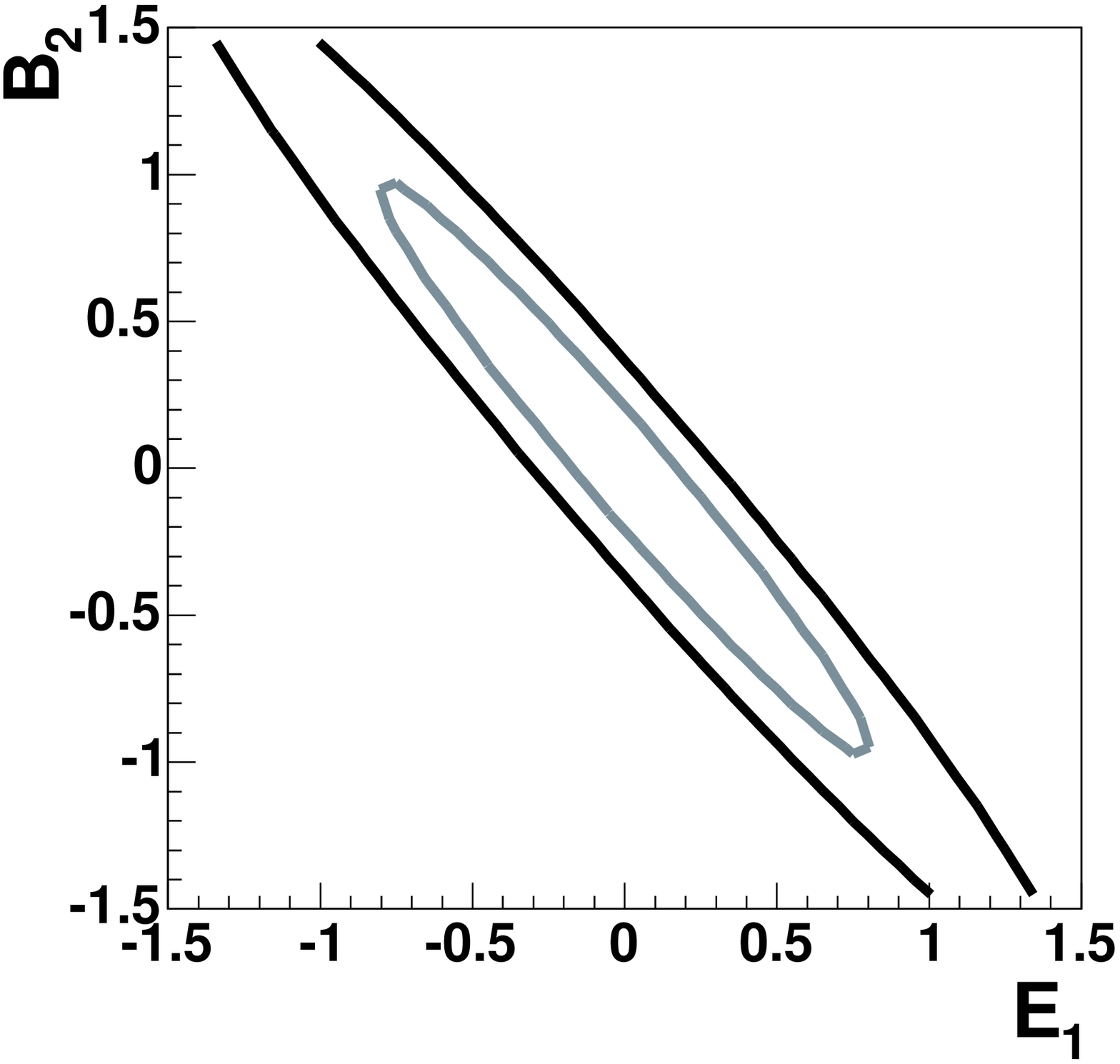}\\
    \includegraphics[width=0.30\textwidth]{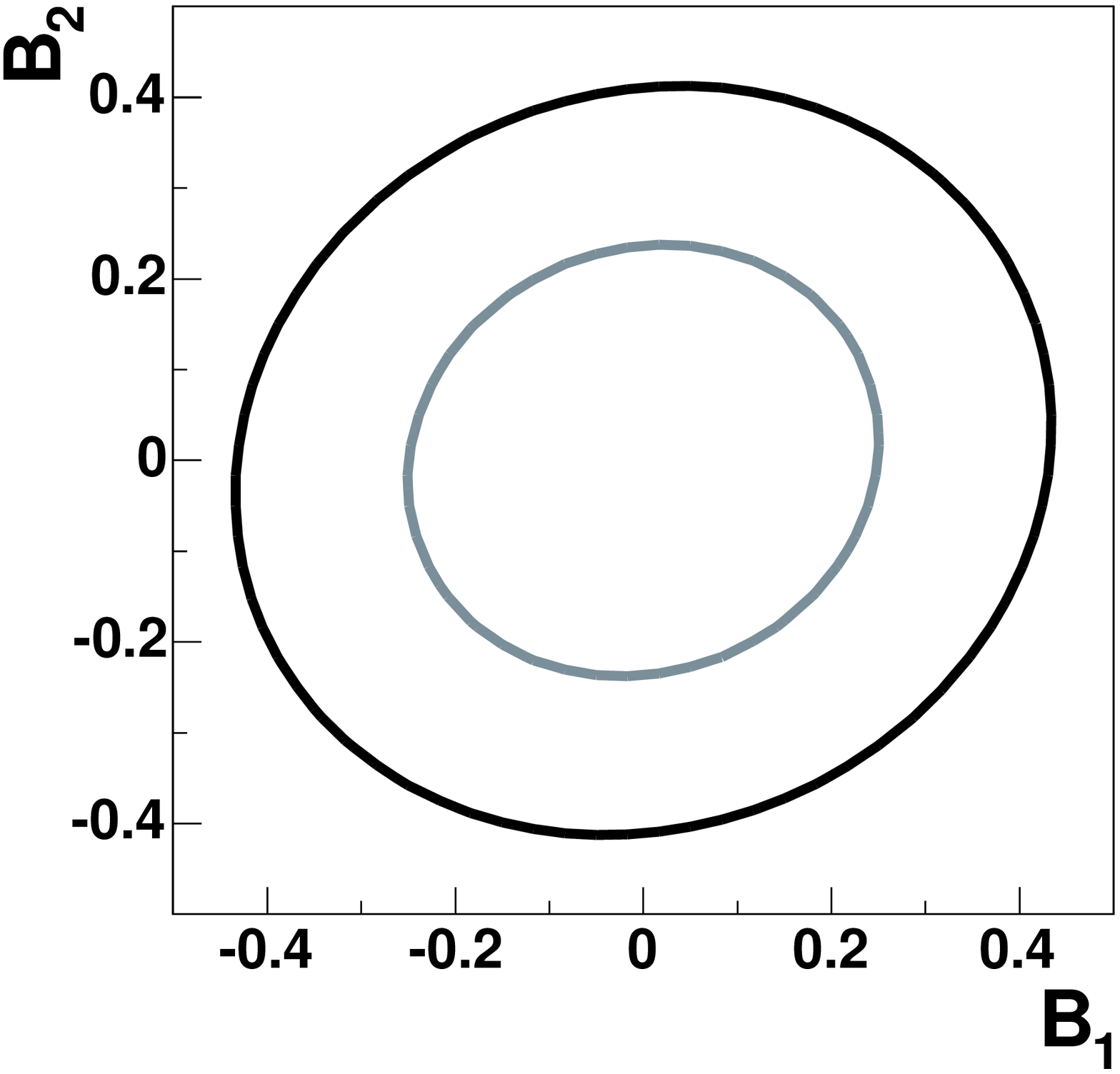}\quad
    \includegraphics[width=0.30\textwidth]{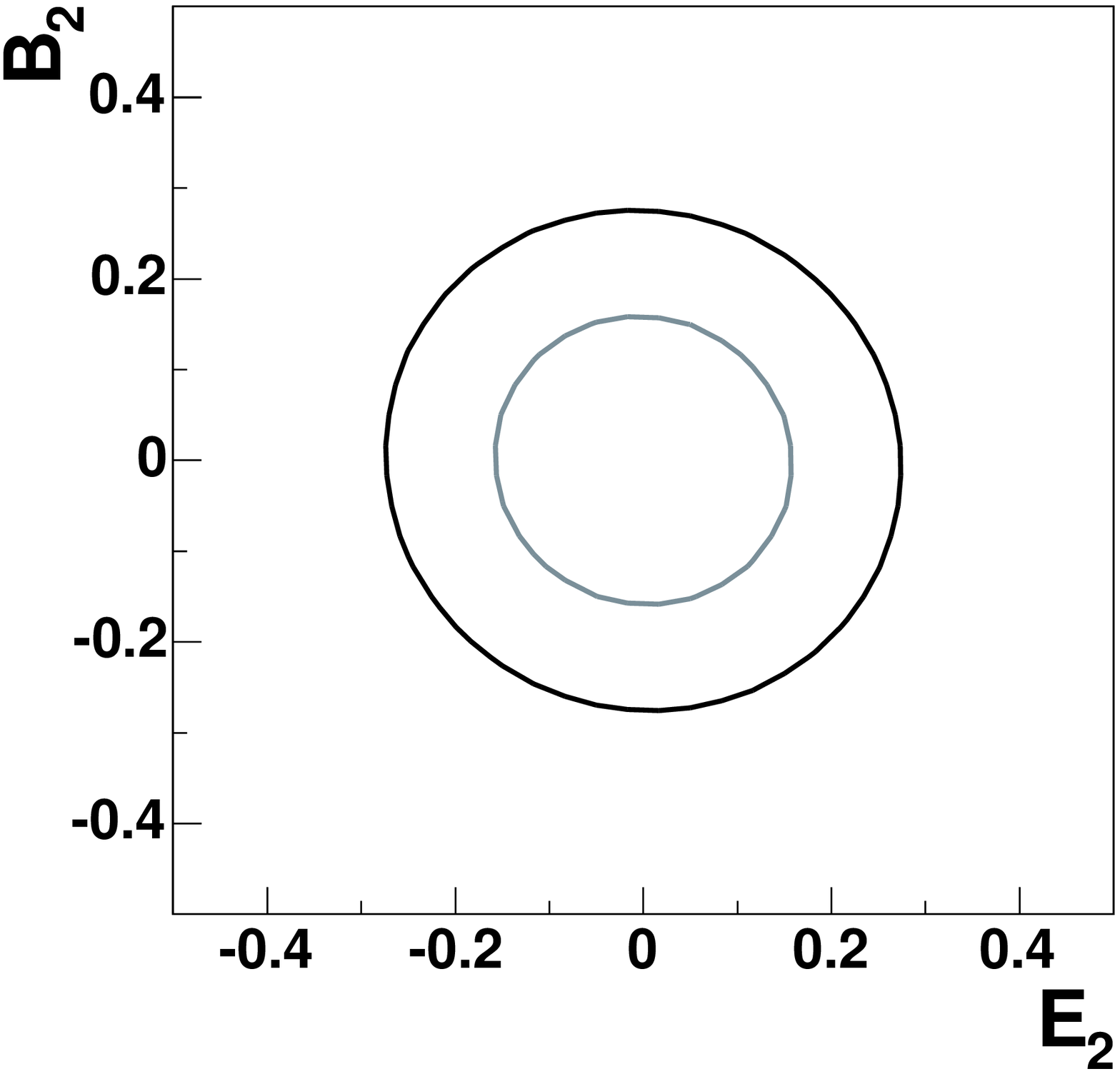}\quad
    \includegraphics[width=0.30\textwidth]{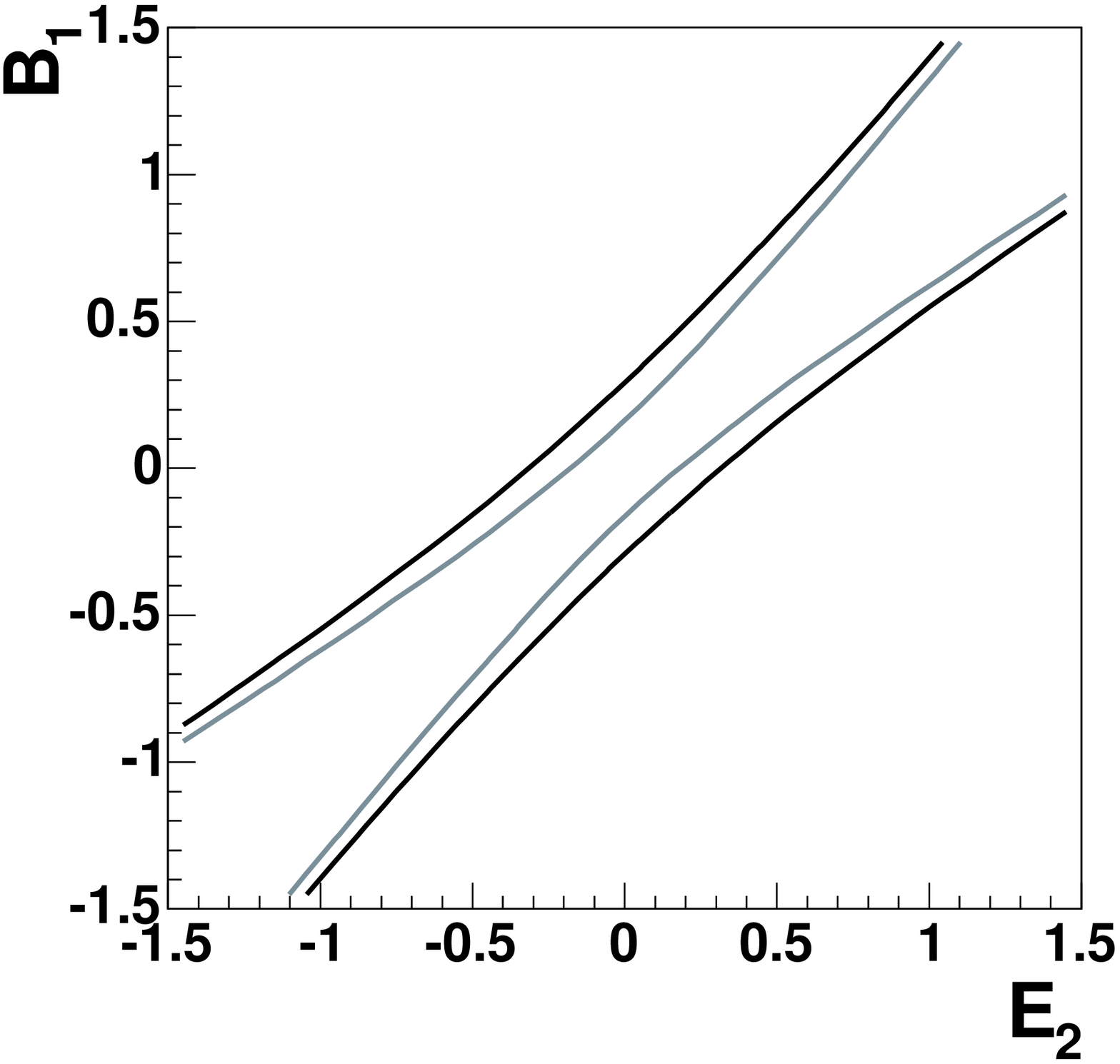}
  \end{center}
  \caption{\label{fig:contours}%
    The $1\sigma$ (dark) and $3\sigma$ (light) exclusion contours
    for~$\Lambda_{\text{NC}}=\unit[500]{GeV}$ and
    $\unit[100]{fb^{-1}}$ at the LHC discussed in the text.}
\end{figure}

For the LHC, the results of the likelihood fits are shown in
fig.~\ref{fig:contours}, setting $\Lambda_{\text{NC}}=\unit[500]{GeV}$
and $K_{Z\gamma\gamma}=K_{ZZ\gamma}=0$.
From the error ellipses in the first and second column of
fig.~\ref{fig:contours}, we derive the following sensitivity limit on
the NC scale for an integrated luminosity of~$100\,\textrm{fb}^{-1}$
(with $100\%$ detection efficiency):
\begin{equation}
  \Lambda_{\mathrm{NC}} \geq \unit[1]{TeV} \;\text{for}\; |\vec E|^2 + |\vec B|^2 = 1\,.
\end{equation}
In the presence of triple gauge couplings among neutral gauge bosons,
these limits do not change significantly.
This analysis indicates that the current collider
bounds~\cite{Abbiendi:2003wv} can be improved at the LHC by an order
of magnitude.

\begin{figure}
  \begin{center}
    \includegraphics[width=0.30\textwidth]{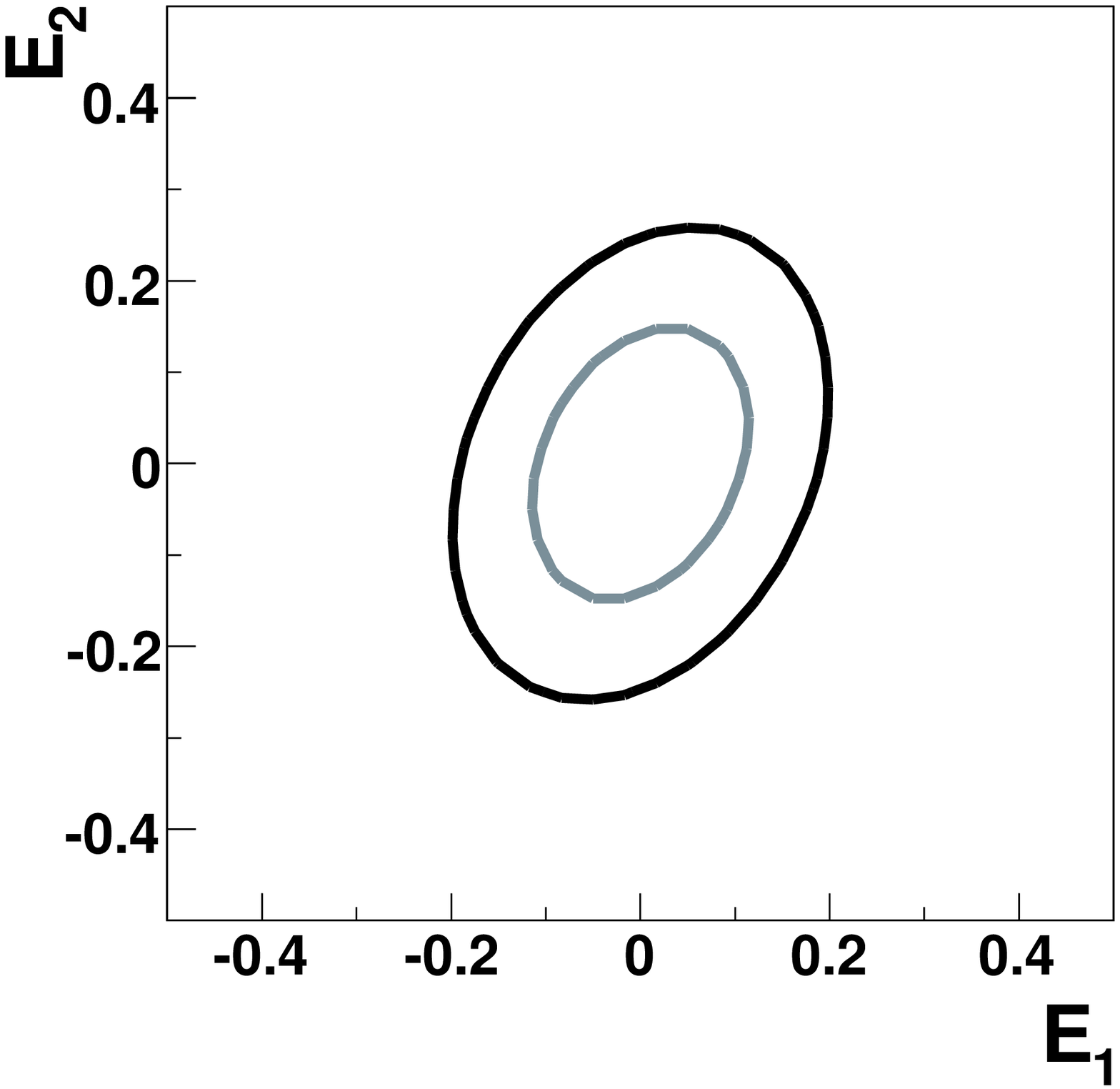}\quad
    \includegraphics[width=0.30\textwidth]{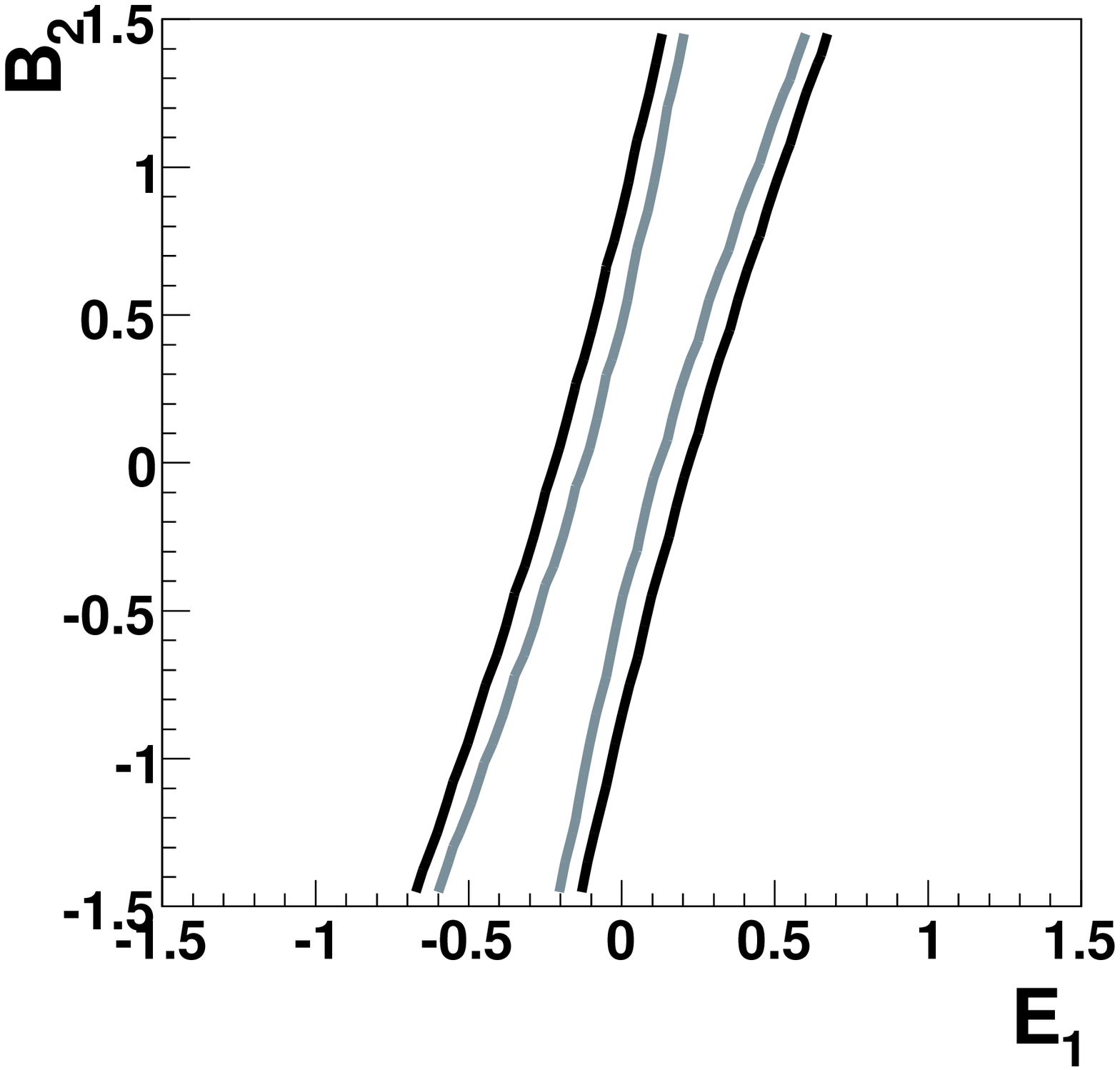}\quad
  \end{center}
  \caption{\label{fig:contours-tevatron}%
    The $1\sigma$ (dark) and $3\sigma$ (light) exclusion contours
    for~$\Lambda_{\text{NC}}=\unit[50]{GeV}$ and $\unit[15]{fb^{-1}}$
    at the Tevatron discussed in the text.}
\end{figure}

A similar analysis of the likelihood fits for the Tevatron, shown in
fig.~\ref{fig:contours-tevatron}, for $\int\mathcal{L} =
\unit[15]{fb^{-1}}$ implies that the sensitivity on the NC scale
reaches $\Lambda_{\textrm{NC}}\sim \unit[130]{GeV}$, which is comparable to
existing LEP bounds~\cite{Abbiendi:2003wv}.

As expected from the Lorentz boost~$(\ref{eq:correlation})$, we find
that in the laboratory frame measurements of~$E_1$ are correlated
with~$B_2$ and measurements of~$E_2$ with~$B_1$. The corresponding
$1\sigma$ and $3\sigma$ contours are depicted for
$\Lambda_{\text{NC}}=\unit[500]{GeV}$ in the
right column of fig.~\ref{fig:contours}.  Since the dependence on~$B_1$
and~$B_2$ in the partonic CMS is very weak, one
expects very elongated ellipses in the laboratory frame, in agreement
with our result.  Due to
statistical fluctuations, the fitted matrix~$X^2$ can have a negative
eigenvalue, as happened in the bottom right plot of
fig.~\ref{fig:contours}.  This sign is
unphysical and as expected changes with the random number sequence
used in the simulations.  For all practical purposes, 
the error ellipses for $(E_1,B_2)$ and $(E_2,B_1)$ should be viewed as
straight lines.

In the other pairs $(E_1,E_2)$, $(B_1,B_2)$ and $(E_i,B_i)$, we find
no correlations.  Indeed, the only source of violation of rotational
invariance is~$\theta^{\mu\nu}$ itself. Therefore no
correlations between~$E_1$ and~$E_2$, as well as between~$B_1$
and~$B_2$, are expected since the measurements are related by a rotation around
the beam axis by $\pi/2$.

Having established the absence of correlations that are not of purely
kinematical origin, one can avoid expensive non-linear
$4$-parameter fits in subsequent work that will take higher orders
in~$\theta^{\mu\nu}$ into
account~\cite{Alboteanu/Ohl/Rueckl:2005:theta2}.

\section{Conclusions and Outlook}
\label{sec:concl}

We have studied the effect of a noncommutative extension of the
SM on the production of neutral vector bosons at the LHC
and the Tevatron.
We have shown that, under conservative assumptions, values for the NC
scale of $\Lambda_{\text{NC}}\gtrsim\unit[1]{TeV}$ can be probed at
the LHC, while the Tevatron is able to confirm existing limits.

The analysis has been performed to first order in~$\theta^{\mu\nu}$.
In this approximation, unphysical cross sections appear for
large~$\sqrt{\hat s}$ and for $\Lambda_{\text{NC}}$ in the region of
sensitivity.  These have been handled by a pragmatic regularization
described in section~\ref{sec:signals} and appropriate kinematical
cuts. On general grounds, one expects the problem to disappear when
higher orders in~$\theta^{\mu\nu}$ are taken into account.
Preliminary results for the second order contributions confirm this
expectation.  We also find that cancellations in the unpolarized cross
sections at $\mathcal{O}(\theta)$ are not effective
in higher orders, allowing also studies of processes with symmetric
final states like $pp\to\gamma\gamma$, one of the most important
channels at the LHC.  Therefore, the derivation of Seiberg-Witten maps
for $\mathrm{SU}(3)_C\otimes \mathrm{SU}(2)_L\otimes \mathrm{U}(1)_Y$
of the corresponding NCSM in higher orders is important for a
comprehensive analysis of the NCSM at the LHC.

The construction of
the NCSM in $\mathcal{O}(\theta^2)$ and further phenomenological
tests, including $\gamma\gamma$ final states, will be presented in
future publications~\cite{Alboteanu/Ohl/Rueckl:2005:theta2}.

\subsection*{Acknowledgments}
This research is supported by Deutsche Forschungsgemeinschaft,
grant RU\,311/1-1, Research Training Group 1147 \textit{Theoretical
Astrophysics and Particle Physics} and by Bundesministerium f\"ur Bildung und Forschung
Germany, grant 05HT4WWA/2.  A.\,A.~gratefully
acknowledges support from Evangelisches Studienwerk e.\,V.~Villigst.


\appendix
\section{Feynman Rules}
\label{app:feynman}

The Feynman rules corresponding to the NCSM Lagrangian~\cite{NCSM}
have been derived in~\cite{Ohl/Reuter:2004:NCPC,NCSM:FeynmanRules}.  Here, we
collect only the neutral current couplings entering our analysis.
Choosing all momenta as incoming and using the shorthand
notation~$p\theta q = p_\mu q_\nu\theta^{\mu\nu}$
and~$p\theta^\nu = p_\mu\theta^{\mu\nu}$, one has
\begin{subequations}
\begin{align}
\label{eq:vertex-ffV}
  \parbox{40mm}{\includegraphics{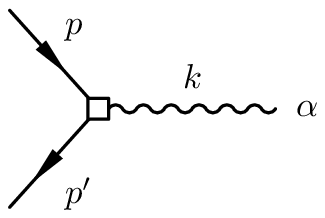}} &
    =   \frac{1}{2} (g_V + g_A\gamma_5)
    \left( k\theta p\gamma^\alpha +
    p\theta^\alpha\fmslash{k} -
    k\theta^\alpha\fmslash{p}\right)\\
\label{eq:vertex-ffVV}
  \parbox{40mm}{\includegraphics{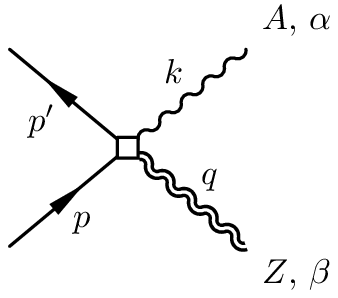}} &
  \begin{aligned}
    = \frac{1}{2}&(g_V + g_A\gamma_5)(g'_V + g'_A\gamma_5) \cdot \\
      &\biggl(  \theta^{\alpha\beta}(\fmslash{q} - \fmslash{k})
               + (k-q)\theta^\beta\gamma^\alpha
               + (q-k)\theta^\alpha\gamma^\beta \biggr)
  \end{aligned}\\
\label{eq:vertex-VVV}
  \parbox{40mm}{\includegraphics{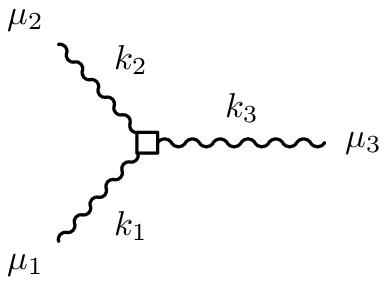}} &
    = \left\{
      \begin{aligned}
        &\; + 2 e \sin(2\theta_W) \,
        K_{Z\gamma\gamma} \cdot V_{\mu_1\mu_2\mu_3}(k_1,k_2,k_3) \\ 
        &\; - 2 e \sin(2\theta_W) \,
        K_{\gamma\gamma\gamma} \cdot V_{\mu_1\mu_2\mu_3}(k_1,k_2,k_3)
      \end{aligned}\right.\,.
\end{align}
\end{subequations}
In the above, the SM vector and axial-vector couplings in the fermion
vertices are given by
\begin{subequations}
\begin{align}
   g_V &= T_3 - 2 Q\sin^2\theta_W\\
   g_A &= T_3\,
\end{align}
while in the $\gamma_{\mu_1}(k_1)\gamma_{\mu_2}(k_2)Z_{\mu_3}(k_3)$ and
$\gamma_{\mu_1}(k_1)Z_{\mu_2}(k_2)Z_{\mu_3}(k_3)$ vertices
\begin{multline}
  V_{\mu_1\mu_2\mu_3}(k_1,k_2,k_3) = \\
        \theta_{\mu_1\mu_2}
          \left[ (k_1 k_3) k_{2,\mu_3} - (k_2 k_3) k_{1,\mu_3} \right]
      + \left( k_1 \theta k_2 \right)
          \left[ k_{3,\mu_1} \eta_{\mu_2\mu_3}
                   - \eta_{\mu_1\mu_3} k_{3,\mu_2} \right] \\
      + \Bigl[   (k_1\theta)_{\mu_1}
                   \left[ k_{2,\mu_3} k_{3,\mu_2} - (k_2 k_3) g_{\mu_2\mu_3} \right]
               - (\mu_1 \leftrightarrow \mu_2)
               - (\mu_1 \leftrightarrow \mu_3)
        \Bigr] \\
  + \text{cyclical permutations of}\;
       \bigl\{ (\mu_1,k_1) , (\mu_2,k_2), (\mu_3,k_3) \, \bigr\}\,.
\end{multline}
\end{subequations}
In the minimal NCSM, the coupling constants~$K_{Z\gamma\gamma}$
and~$K_{ZZ\gamma}$ vanish.  In the non-minimal NCSM, they can take the
values shown in fig.~\ref{fig:K_Zgg/K_ZZg} (see~\cite{NCSM}).

\section{Cross Section $q\bar q\to Z\gamma$}
\label{app:xsect}

In order to be able to express the partonic cross section for $q\bar
q\to Z\gamma$ compactly, we introduce some further abbreviations. For
contractions with the totally antisymmetric
$\epsilon_{\mu\nu\rho\sigma}$ tensor we use the notation
\begin{subequations}
\begin{align}
  \epsilon_{\mu\nu\rho\sigma}p^\mu k^\nu q^\rho r^\sigma
    &= \langle pkqr\rangle \\
  \epsilon_{\mu\nu\rho\sigma} p^\mu k^\nu q^\rho r_\alpha \theta^{\alpha\sigma}
    &= \epsilon_{\mu\nu\rho\sigma}p^\mu k^\nu q^\rho r\theta^\sigma
     = \langle pkqr\theta\rangle \\
  \theta^{\mu\nu}\epsilon_{\mu\nu\rho\sigma}
    &= \langle \theta \rangle_{\rho\sigma} \\
  \theta^{\mu\nu}\epsilon_{\mu\nu\rho\sigma} p^\rho k^\sigma
    &= \langle \theta \rangle_{\rho\sigma}p^\rho k^\sigma = 
        p\langle \theta \rangle q
\end{align}
\end{subequations}
The squared amplitude
\begin{multline}
 |A|^2
   = (|A|^2)^{\text{SM}}
      + (|A|^2)^{\text{NC}}_{tu}
      + (|A|^2)^{\text{NC}}_{tc+uc}
      + (|A|^2)^{\text{NC}}_{t\gamma + u\gamma}
      + (|A|^2)^{\text{NC}}_{tZ + uZ}
      + \mathcal{O}(\theta^2)
\end{multline}
is given by the SM contribution
\begin{subequations}
\begin{equation}
  (|A|^2)^{\text{SM}}
     = \frac{8\pi^2}{3}
         \left( \frac{Q_q\alpha}{\sin{\theta_W}\cos{\theta_W}}\right)^2
                (g_V^2 + g_A^2)\left(\frac{u}{t} + \frac{t}{u} 
                    + \frac{2 m_Z^2 s}{tu} \right)
\end{equation}
and SM/NCSM interference terms
\begin{multline}
  (|A|^2)^{\text{NC}}_{tu} =
         2\mathrm{Re} \left(   A_t^{\text{SM}}A_u^{*\,\text{NC}}
                             + A_t^{\text{NC}}A_u^{*\,\text{SM}} \right) = \\
     \frac{2\pi^2}{3}
        \left( \frac{Q_q\alpha}{\sin{\theta_W}\cos{\theta_W}} \right)^2 (2g_Vg_A)
        \left(   8\langle p_1k_1k_2k_1\theta\rangle
               + 8\langle p_1k_1k_2k_2\theta\rangle \right)
        \left(   \frac{1}{u} - \frac{1}{t}\right)\,,
\end{multline}
\begin{multline}
  (|A|^2)^{\text{NC}}_{tc+uc} =
         2\mathrm{Re} \left(   A_t^{\text{SM}}A_c^{*\,\text{NC}} 
                             + A_u^{\text{SM}}A_c^{*\,\text{NC}} \right) = \\
     -\frac{2\pi^2}{3}
       \left( \frac{Q_q\alpha} {\sin{\theta_W}\cos{\theta_W}} \right)^2 (2g_Vg_A)
       \left[   2k_1\langle\theta\rangle k_2 \frac{m_Z^2}{t}
              + 2p_1\langle\theta\rangle k_2
                   \left( \frac{s+t}{u} - \frac{s+u}{t} \right) \right]\,,
\end{multline}
\begin{multline}
  (|A|^2)^{\text{NC}}_{t\gamma + u\gamma} =
         2\mathrm{Re} \left(   A_t^{\text{SM}}A_{\gamma}^{*\,\text{NC}}
                             + A_u^{\text{SM}}A_{\gamma}^{*\,\text{NC}} \right) \\
     = \frac{16\pi^2}{3}
          Q_q^2\alpha^2 K_{Z\gamma\gamma} g_A\frac{1}{s} T(p_1,p_2,k_1,k_2)\,,
\end{multline}
\begin{multline}
  (|A|^2)^{\text{NC}}_{tZ + uZ} =
         2\mathrm{Re} \left(   A_t^{\text{SM}}A_{Z}^{*\,\text{NC}}
                             + A_u^{\text{SM}}A_{Z}^{*\,\text{NC}} \right) \\
     = \frac{8\pi^2}{3}
          \frac{Q_q\alpha^2}{\sin{\theta_W}\cos{\theta_W}} K_{ZZ\gamma}
          (2g_Vg_A) \frac{1}{s-m_Z^2+\ii m_Z\Gamma_Z} T(p_1,p_2,k_1,k_2)\,.
\end{multline}
\end{subequations}
Note that the $s$-channel contributions
differ only by the coupling constants and the propagator
because the triple gauge couplings contribute the same factor
\begin{multline}
  T(p_1,p_2,k_1,k_2) = \\
    \left[  p_1\langle\theta\rangle k_1
              \left( -2s \right)
          + p_1\langle\theta\rangle k_2
              \left( u - t - 4s + \frac{u(s+u)}{t} - \frac{3st+2s^2+t^2}{u} \right) \right. \\
          + k_1\langle\theta\rangle k_2 m_Z^2 \left( 1 - \frac{u}{t} \right)
          - 4\langle p_1k_1k_2p_1\theta\rangle
              \left( 1 + \frac{t+2s}{u} + \frac{u}{t} \right) \\
          + \left.
            4\langle p_1k_1k_2k_1\theta\rangle
              \left( 1 + 3\frac{u}{t} - 2\frac{t}{u} \right)
          + 4\langle p_1k_1k_2k_2\theta\rangle
              \left( 1 + \frac{t+4s}{u} - 2\frac{s}{t} \right) \right]\,.
\end{multline}
This analytical result has been verified numerically by comparing with the
scattering amplitudes calculated with the help of an
unpublished extension of the matrix element generator
O'Mega~\cite{Ohl/etal:Omega}.  The latter has also been employed for the
calculation of the complete $q\bar q\to\ell^+\ell^-\gamma$ scattering
amplitude used in the simulations.

\end{document}